\pdfoutput=1

\documentclass[11pt]{article}

\usepackage[preprint]{acl}

\usepackage{times}
\usepackage{latexsym}

\usepackage[T1]{fontenc}

\usepackage[utf8]{inputenc}

\usepackage{microtype}

\usepackage{inconsolata}

\usepackage{graphicx}

\usepackage{url}
\title{SpeechAccentLLM: A Unified Framework for Foreign Accent Conversion and Text to Speech}

\author{
  \textbf{Zhuangfei Cheng\textsuperscript{1}},
  \textbf{Guangyan Zhang\textsuperscript{2}},
  \textbf{Zehai Tu\textsuperscript{2}},
  \textbf{Yangyang Song\textsuperscript{1}},
\\
  \textbf{Shuiyang Mao\textsuperscript{3}},
  \textbf{Xiaoqi Jiao\textsuperscript{2}},
  \textbf{Jingyu Li\textsuperscript{2}},
  \textbf{Yiwen Guo\textsuperscript{4}},
  \textbf{Jiasong Wu\textsuperscript{1}},
\\
  \textsuperscript{1}Southeast University,
  \textsuperscript{2}LIGHTSPEED,
  \textsuperscript{3}Tencent Hunyuan,
  \textsuperscript{4}Independent Researcher,
\\
  \small{
    \textbf{Correspondence:} \href{mailto:jswu@seu.edu.cn}{jswu@seu.edu.cn}
  }
}

\begin{document}
\maketitle
\begin{abstract}
    Foreign accent conversion (FAC) in speech processing remains a challenging task. Building on the remarkable success of large language models (LLMs) in Text-to-Speech (TTS) tasks, this study investigates the adaptation of LLM-based techniques for FAC, which we term SpeechAccentLLM. At the core of this framework, we introduce SpeechCodeVAE, the first model to integrate connectionist temporal classification (CTC) directly into codebook discretization for speech content tokenization. This novel architecture generates tokens with a unique "locality" property, as validated by experiments demonstrating optimal trade-offs among content faithfulness, temporal coherence, and structural recoverability. Then, to address data scarcity for the FAC module, we adopted a multitask learning strategy that jointly trains the FAC and TTS modules. Beyond mitigating data limitations, this approach yielded accelerated convergence and superior speech quality compared to standalone FAC training. Moreover, leveraging the salient properties of our discrete speech representations, we introduce SpeechRestorer, a postprocessing architecture designed to refine LLM-generated outputs. This module effectively mitigates stochastic errors prevalent in LLM inference pipelines while enhancing prosodic continuity, as validated by ablation experiments.
\end{abstract}

\section{Introduction}
The goal of foreign accent conversion is to modify nonnative accents in second language (L2) speech while preserving linguistic content and speaker identity. This task significantly benefits language education and cross-cultural communication by reducing barriers to speech intelligibility. 
Accent conversion (AC) involves transforming speech with a source accent into a target accent. FAC, as a subtask of AC, differs fundamentally in its scope of operation: FAC exclusively transforms nonnative accented speech into native accented speech, while AC enables bidirectional conversion (including native-to-nonnative transformation) and cross-accent conversion between arbitrary accents. FAC and AC are challenging tasks because they require not only changes in pronunciation but also modifications in prosody and phoneme duration. Additionally, the shortage of accented datasets is another common difficulty in both FAC and AC tasks. 

The early approaches for FAC were similar to the process of voice conversion (VC), which required reference first language (L1) speech data to eliminate the nonnative accents in the generated speech~\cite{refer_native}.  As the paired L1 data were difficult to obtain,  these methods might be ineffective during inference. Later, methods that did not require referencing L1 were developed~\cite{AC_End-to-end,Zero-shotFAC,converting-FAC}. 
Some of these approaches used paired L1 data to train the AC model ~\cite{converting-FAC}. Meanwhile, some methods were developed, which did not require supervised data ~\cite{AC_usetts}. Thanks to recent advancements in TTS and VC technologies~\cite{cosyvoice,cosyvoice2,xtts,zuo2025enhancing}, high-quality synthesized paired native accented speech can be easily generated and utilized in FAC. The focus of the study in AC gradually shifted to the modeling design. Recent works include flow-based models~\cite{AC_flow} and TTS-guided frameworks~\cite{AC_usetts} to improve modeling in AC tasks. Despite progress, these methods still have limitations in handling speaker-dependent accent variations and require substantial training data. 

Recent advancements in LLMs have demonstrated their remarkable potential in speech processing, particularly in speech synthesis. A pivotal factor driving this success lies in the capability of discrete speech representations to effectively capture relevant speech characteristics. Current discrete speech representations can be broadly categorized into two types: semantic discrete representations and acoustic discrete representations.

Semantic discrete representations preserve linguistic content and exhibit strong correlations with phonemes. Early approaches typically employed K-means clustering on pre-trained speech encoders (e.g., Hubert~\cite{hubert}, Data2vec~\cite{data2vec}) to derive these representations, which were subsequently applied to downstream tasks. For example, SpeechGPT~\cite{speechgpt} for speech recognition by integrating semantic tokens into GPT-style architectures and Polyvoice~\cite{polyvoice} for speech translation tasks. However, K-means quantization inherently discards substantial speech information, leading to performance degradation in downstream applications. Recent innovations, such as Cosyvoice~\cite{cosyvoice}, have adopted supervised training to obtain semantic tokens combined with flow matching techniques, achieving state-of-the-art performance in Mandarin TTS synthesis. In contrast, discrete acoustic representations encode comprehensive acoustic information from speech signals. Although these representations enable complex generation tasks such as audio/music synthesis (e.g., AudioLM~\cite{audiolm}), they introduce significant technical challenges: the bitrate of discrete sequences exceeds the input capacity of conventional LLMs, and the cardinality of the codebook substantially exceeds typical vocabulary sizes, leading to increased training complexity and computational overhead.

Inspired by the successful adaptation of LLMs in speech processing, we identify the unique potential of semantic discrete representations for FAC tasks. These representations inherently exclude timbre information while effectively encoding semantic attributes, making them particularly suited for addressing FAC's core challenge: decoupling and transferring accent features without speaker identity interference. This observation motivates our proposal of SpeechAccentLLM, the first unified LLM-based framework that jointly optimizes TTS and FAC through three key innovations,
\begin{enumerate}
\item a CTC-regularized SpeechCodeVAE that extracts speaker-agnostic speech content tokens with superior locality.
\item a unified framework for FAC and TTS that leverages TTS data to compensate for FAC data scarcity.
\item a SpeechRestorer module that refines LLM outputs through token-level error correction.
\end{enumerate}
\par

\begin{figure}[t]
\centering

\includegraphics[width=0.66\columnwidth]{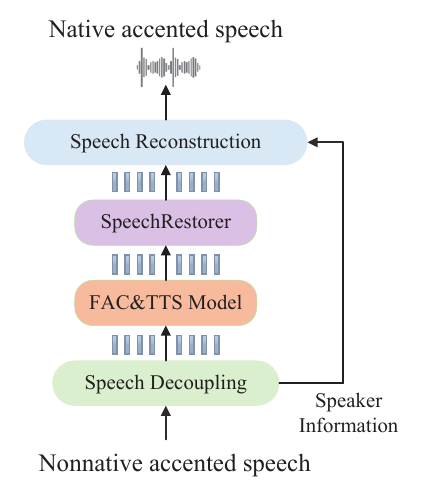} 
\caption{The overview of SpeechAccentLLM for FAC inference. Speech Decoupling extracts nonnative accented content tokens and speaker information from nonnative accented speech. The FAC\&TTS Model takes nonnative accented content tokens as input and outputs predicted native accented content tokens. SpeechRestorer post-processes the speech content tokens predicted by the FAC\&TTS Model. Finally, Speech Reconstruction combines the output content tokens from SpeechRestorer with speaker information to reconstruct native accented speech.}
\label{Fig: fig1}
\end{figure}
The FAC inference architecture of our proposed SpeechAccentLLM framework is presented in ~\autoref{Fig: fig1}, which includes four parts: the Speech Decoupling and Speech Reconstruction stages of SpeechCodeVAE, the FAC\&TTS Model, and the SpeechRestorer. We note the absence of phoneme information in accented speech, while conventional phoneme-based systems~\cite{AC_usetts,vits} are prone to error propagation when processing accented speech. This is because the rule-based text-to-phoneme conversion in the frontend is typically limited and may fail for low-resource languages. SpeechCodeVAE is designed to extract and reconstruct high-quality speech content tokens for arbitrary speech. It comprises the Speech Decoupling stage and Speech Reconstruction stage. The Speech Decoupling stage leverages the features of Whisper encoder~\cite{whisper}, enhanced via CTC-guided vector quantization to extract discrete content tokens. These discrete content representations are highly correlated with phonetic information in speech, enabling stable training for FAC without explicit phonetic annotations. The reconstruction stage employs a prosody adapter and a VITS-based backend to reconstruct content and timbre representations into the final synthesized speech.

To tackle the challenge of limited dataset availability for the FAC task, we adopted a unified training framework. Within a multitask learning paradigm, we use the rich data from the TTS task to inform the FAC task, thereby facilitating faster training convergence and enhancing conversion performance. In addition to addressing the critical challenge of spurious inference errors inherent in LLM-based speech synthesis, we introduce SpeechRestorer: a novel module inspired by BERT's token restoration mechanism for masked language modeling~\cite{bert}. Operating on localized speech content tokens, SpeechRestorer effectively detects and rectifies inconsistencies arising from LLM-generated outputs, thereby enhancing speech fluency and optimizing performance.

\section{SpeechCodeVAE}
\label{sec:speechtokenizer}
\begin{figure}[t]
\centering

\includegraphics[width=\columnwidth]{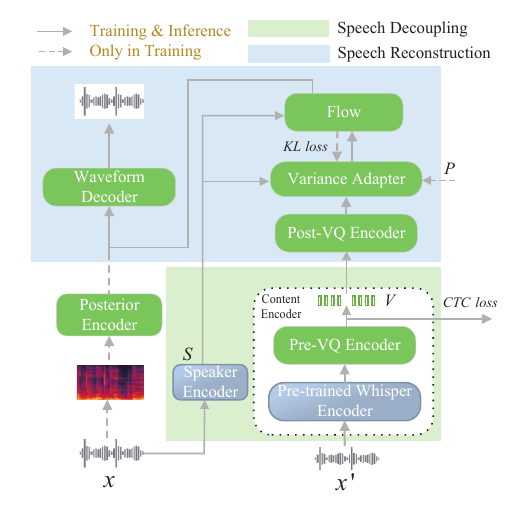} 
\caption{The overview of SpeechCodeVAE. The Speech Decoupling stage extracts speech content tokens $\mathbf{V}$ and speaker timbre embeddings $\mathbf{S}$, and the Speech Reconstruction stage reconstructs the decoupled information. }
\label{Fig: fig2}
\end{figure}

The SpeechCodeVAE model aims to decouple and reconstruct speech signals through three disentangled representations: speech content tokens $\mathbf{V}$, speaker timbre embeddings $\mathbf{S}$, and prosodic features $\mathbf{P}$. The prosodic representation $\mathbf{P}$ is exclusively utilized during the training phase, while its counterpart is automatically synthesized by the model during the inference phase. As shown in~\autoref{Fig: fig2}, our architecture incorporates four specialized modules in addition to the inherent modules of VITS~\cite{vits}: Speaker Encoder, Content Encoder, Post-VQ Encoder, Variance Adapter. SpeechCodeVAE comprises two core stages: Speech Decoupling and Speech Reconstruction. The Speech Decoupling stage includes Content Encoder and Speaker Encoder, which are responsible for extracting $\mathbf{V}$ and $\mathbf{S}$ respectively. The Speech Reconstruction stage integrates four components: Post-VQ Encoder, Variance Adapter, Flow module, and Waveform Decoder, which collectively rely on the extracted $\mathbf{V}$ and $\mathbf{S}$ to reconstruct the speech waveform. At this stage, we also integrate the prosodic representation $\mathbf{P}$ into the model.

The functions of the model modules are as follows: the perturbation of original waveforms $x$ into distorted waveforms $x'$, the Speaker Encoder extracting $\mathbf{S}$ from raw speech $x$, the Content Encoder generating $\mathbf{V}$ from perturbed input $x'$, the Post-VQ Encoder reconstructing content features from $\mathbf{V}$, the Variance Adapter fusing $\mathbf{V}$ with prosodic information $\mathbf{P}$, the Flow module injecting $\mathbf{S}$, the Posterior Encoder presenting posterior probability, and the Waveform Decoder synthesizing final speech. The training process incorporates adversarial guidance through a discriminator, ensuring high-fidelity waveform generation while maintaining efficient content-speaker-prosody disentanglement.

\par

\textbf{Content Encoder}
Our Content Encoder aims to discretize speech content into $\mathbf{V}$ while ensuring these representations exhibit strong robustness. The Content Encoder comprises three components: 1) Pretrained Whisper Encoder, 2) Pre-VQ Encoder, and 3) Vector Quantization (VQ) module.

In the model training, we take $x'$ as input to the Content Encoder for extracting $\mathbf{V}$. This approach is adopted because modifying the $f_0$ or formant positions alters the timbre characteristics in speech signals, which not only helps mitigate timbre leakage issues but also facilitates the reconstruction of $\mathbf{S}$ information. The perturbation methods include: $x$ undergoes controlled perturbations through the NANSY module~\cite{nansy}, randomly modifying fundamental frequency ($f_0 \pm 20\%$) and formant positions ($\pm15\%$) across 50\% of training samples.

We employ the Pretrained Whisper Encoder because Whisper, as a multilingual ASR model, effectively removes speaker-specific information from speech through its pre-trained encoder~\cite{cosyvoice}. In contrast, other speech self-supervised learning (SSL) models (e.g., Hubert~\cite{hubert}, WavLM~\cite{wavlm}) retain comprehensive speech information in their features, which hinders content-specific feature extraction.

The Pre-VQ Encoder adopts a hybrid architecture combining multiple convolutional layers with transformer encoder layers. This design facilitates dimensionality reduction and further extracts content features using CTC loss. We use international phonetic alphabet (IPA) for the labels of CTC.

The Vector Quantization (VQ) module is responsible for discretizing and quantizing speech content information, employing an Exponential Moving Average (EMA) mechanism for parameter updates during training.

\textbf{Post-VQ Encoder}
This module is designed to reconstruct the loss associated with speech content discretization. We therefore define \(L_{VQ}\) as the L2 loss between the continuous representations that are the outputs of the Pre-VQ Encoder and the outputs of the Post-VQ Encoder.

\textbf{Variance Adapter}
The Variance Adapter serves two purposes: integrating prosodic features $\mathbf{P}$ into content representations during training, and predicting $\mathbf{P}$ during inference. We employ $f_0$ as the acoustic correlate of $\mathbf{P}$. The Variance Adapter comprises an $f_0$ predictor and an encoder module.

The $f_0$ predictor, composed of 1D convolutional layers and transformer encoder blocks, is designed to capture prosodic features by jointly modeling speaker and content representations. It takes the Post-VQ Encoder outputs and speaker embeddings $\mathbf{S}$ as inputs, generating normalized $f_0$ values, whereby normalization mitigates pitch variations across different voice timbres and facilitates model training stability. During training, $f_0$ is directly extracted from the original waveforms $x$ and the gradient of $f_0$ predictor is detached. During inference, $f_0$ is predicted by the $f_0$ predictor and then combined with the content representations through the encoder module. 

Subsequently, the encoder module integrates these normalized $f_0$ features with content representations through a dual-stream architecture: (1) the $f_0$ values are discretized and aligned with content vectors via embedding layers to bridge their latent spaces, and (2) the original Post-VQ Encoder outputs are processed in parallel. These two streams are jointly encoded by transformer layers, where the discretization of $f_0$ explicitly enables learnable alignment between prosodic patterns and linguistic content, thereby achieving unified feature fusion for downstream waveform reconstruction.

The Speaker Encoder employs a pre-trained speaker recognition model to extract speaker timbre embeddings $\mathbf{S}$. The remainder of the architectural components maintain identical configuration to VITS~\cite{vits}. Following VITS, our model consists of a generator and a discriminator. The generator loss is described as,
\[L=L_{VQ}+L_{CTC}+L_{f_0}+L_{mel}+L_{KL}+L_{adv}+L_{fm}\]
Here, \(L_{VQ}\) represents the reconstruction loss of the Post-VQ Encoder. \(L_{CTC}\) represents the CTC loss of the Content Encoder. \(L_{f_0}\) represents the $f_0$ predictor loss of the Variance Adapter. The training objective additionally incorporates mel reconstruction loss\(L_{mel}\), KL divergence loss\(L_{KL}\), adversarial loss\(L_{adv}\) and feature-matching loss\(L_{fm}\) components inherited from the VITS~\cite{vits} framework.

\section{Unified framework for FAC and TTS}
\label{sec:framework}

\begin{figure}[t]
\centering
\includegraphics[width=0.96\linewidth]{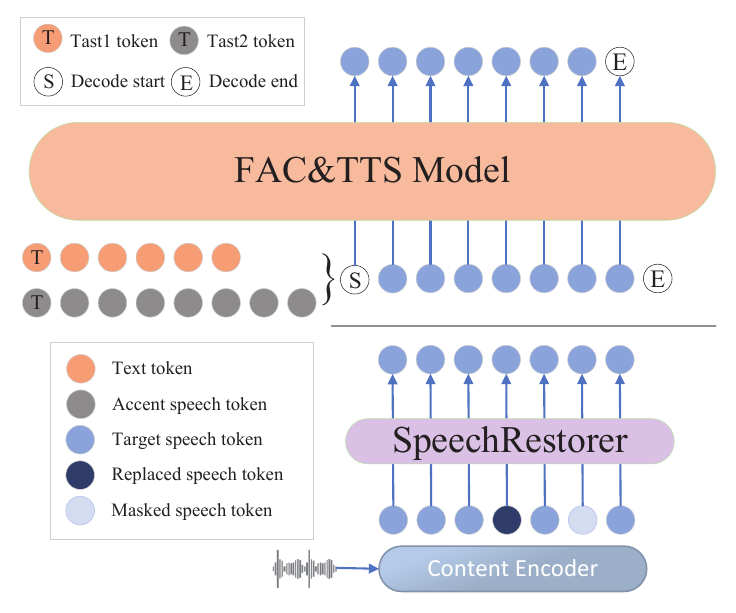}
\caption{Training processes of the FAC\&TTS Model and SpeechRestorer: the two modules are trained separately, all speech is converted into speech content tokens via the Content Encoder for model training.}
\label{Fig: fig3}
\end{figure}

With the speech content tokens obtained by the SpeechCodeVAE, we can directly utilize paired nonnative-native accented data to train a model for the FAC task. However, since the nonnative accented dataset is small-scale and prone to convergence challenges during training, we integrate the TTS task as an auxiliary objective for joint training. Additionally, to improve the instability of language model inference and fully leverage the locality of speech content tokens, we propose the SpeechRestorer, which is designed to filter the outputs of the FAC\&TTS Model, mitigating random errors while enhancing the fluency of synthesized speech. The training processes of FAC\&TTS Model and SpeechRestorer are shown in~\autoref{Fig: fig3}.

Nonnative accented speech datasets often lack corresponding native accented counterparts, primarily due to the inherent difficulty L2 speakers face in producing canonical L1 pronunciations. To address this, we leverage the pre-trained TTS model to generate native accented counterparts for the accented dataset.  

We implemented a transformer decoder-based LLM, designated as the FAC\&TTS Model, which served as the core architecture for joint training of FAC and TTS systems. During the training of FAC\&TTS Model, all speech inputs are first converted into speech content tokens via SpeechCodeVAE. As a multi-task model, FAC\&TTS Model employs Task ID as the initial token to specify the target task. 
For the TTS task: the input sequence begins with text tokens, followed by a decode start token to trigger autoregressive generation. The model then predicts the corresponding speech content tokens sequentially, terminating with a decode end token.     
For the FAC task: the input sequence starts with nonnative accented speech content tokens, with the decode start token initiating the decoding process. The objective is to generate native accented speech content tokens, concluding with the decode end token.

The training of SpeechRestorer is decoupled from FAC\&TTS Model. Specifically, the target speech tokens from the FAC\&TTS Model are employed as training inputs for SpeechRestorer.
During training, 10\% of the tokens are randomly replaced and another 10\% are masked, with the objective of training SpeechRestorer to recover the original tokens. During inference, SpeechRestorer filters and refines the outputs of FAC\&TTS Model before synthesizing the final speech waveform.

\section{Experiment Settings}
\label{sec:exps}

\subsection{Datasets}
Our framework utilizes two primary datasets with distinct purposes:

\textbf{Multilingual Base Corpus}:
We construct this corpus using three phonetically diverse languages providing coverage for 92\% of IPA phonemes:
\begin{itemize} 
\item Chinese: AISHELL-1~\cite{aishell} (180hrs, 400 speakers) 
\item English: LibriSpeech train-clean-360~\cite{librispeech} (360hrs, 921 speakers) 
\item Japanese: JVS corpus~\cite{jvs} (22hrs, normal speech subset, 100 speakers) 
\end{itemize}

The combined corpus contains approximately 250,000 speech segments (6s average duration) totaling 562 hours. All audio was processed using Sox with: 16 kHz resampling, -27 dB loudness normalization.

\textbf{FAC\&TTS Corpus}:
For the FAC, we use the L2-ARCTIC corpus which is the subset of FAC task baseline~\cite{Zero-shotFAC} corpus, excluding the ARC-TIC dataset~\cite{cmu}. Similarly to the baseline, we excluded four speakers from L2-ARCTIC (NJS, YKWK, TXHC, and ZHAA) for the test set. The final corpus comprised 20 nonnative English speakers from six L1 backgrounds (Arabic, Hindi, Korean, Mandarin, Spanish, and Vietnamese) and contains 9.3 hours of speech, with each speaker contributing about 1000 utterances.

To create native accented data corresponding to the nonnative accented dataset, we selected the VITS~\cite{vits} model trained on the LJSpeech dataset as the native accented speech TTS model. The VITS model is a parallel TTS system that demonstrated excellent stability and accuracy. Although the speech synthesized by VITS is of a single speaker, our SpeechCodeVAE model exclusively extracts its content information without being influenced by speaker information, thus this configuration fully satisfied our requirements.

For the TTS, we select speech data from the LibriSpeech train-clean-360 corpus and the speech generated by the above-mentioned VITS model, resulting in a total of 370 hours of data. To address the data imbalance between the FAC and TTS tasks, we employ oversampling to balance the datasets between two tasks.

\subsection{Implementation Details}

SpeechCodeVAE is trained on the \textbf{Multilingual Base Corpus} to obtain general representations. Specifically, the Content Encoder uses Whisper-medium (50Hz, frozen); Speaker Encoder adopts ECAPA-TDNN~\cite{xvector} (512-dim, frozen); VQ layer with 1024 codebook entries.

FAC\&TTS Model, trained on the \textbf{FAC\&TTS Corpus} for FAC and TTS tasks, is an 8-layer transformer decoder (512-dim, 8 heads). We maintain a dropout rate of 0.1. The data ratio for AC and TTS tasks is 1:1 during training. 

SpeechRestorer is also trained on the \textbf{FAC\&TTS Corpus}. The architecture of SpeechRestorer aligns with the standard BERT framework~\cite{bert}, implemented as a 4-layer bidirectional transformer encoder with 512-dimensional hidden states and 4 attention heads per layer. We maintain a consistent dropout rate of 0.1.

For all our models, the batch size is 32 with mixed precision, the ratio of the training set to the validation set during training is 9:1.

\subsection{Evaluation Metrics} \label{subsec:metrics}

We establish a dual evaluation protocol that combines perceptual assessments of human listeners with quantitative acoustic analyses. For subjective evaluation, 20 native English speakers participate in controlled listening tests, assessing three key perceptual dimensions: speech naturalness (5-point CMOS scale), accentedness (10-point expert rating) which refers to the degree of nonnative phonetic characteristics present in speech, and speaker similarity (5-point triplet test SMOS scale). CMOS and SMOS were conducted using reference anchors to minimize individual bias.

Objective evaluation employs five complementary metrics: word error rate (WER) measuring content preservation using the Whisper ASR model~\cite{whisper}; speaker similarity (Sim-O/R) quantifies speaker resemblance, where Sim-O measures similarity between synthesized speech and the reference speaker, while Sim-R evaluates similarity between synthesized outputs and reconstructed reference speech. This metric is computed by extracting speaker embeddings through a pre-trained speaker verification model\footnote{\url{https://github.com/microsoft/UniSpeech/tree/main/downstreams/speaker\_verification}}, followed by cosine similarity calculation between embedding pairs; two metrics (De-duplication Efficiency and Speed Robustness) are introduced to evaluate speech discrete tokens, De-duplication Efficiency~\cite{stab} quantifies temporal compression through consecutive tokens redundancy analysis, and Speed Robustness~\cite{stab} assesses speaking rate invariance via quantifying variation in discrete tokens under double-speed speech conditions. 

\section{Experiments Results}
\label{sec:results}

\subsection{Foreign Accent Conversion Analysis}
We selected 100 utterances from the test set of \textbf{Accent Conversion Corpus} to assess FAC performance. We adopted the framework proposed in~\cite{Zero-shotFAC} as our baseline model. As shown in ~\autoref{tab:fac}, SpeechAccentLLM demonstrates significant accent reduction across four L1 backgrounds. The 25\% improvement in accentedness score (1.86 vs baseline 2.48) stems from the assistance of the text information during the training process. The WER reduction from 14.4\% to 9.1\% confirms improved intelligibility without compromising linguistic content. Interestingly, we also found that CMOS values and listeners' accent strength ratings showed a strong negative correlation (r = -0.82), suggesting that perceived naturalness was directly related to accentedness.

\begin{table}[th]
\centering
\caption{The FAC Performance Evaluation}
\label{tab:fac}
\vspace{4pt}
\scalebox{0.77}{\begin{tabular}{ccccc}
\hline
&Sim-O $\uparrow$&WER$\downarrow$&Accentedness$\downarrow$&CMOS$\uparrow$  \\
\hline
Baseline&0.558&0.144 &2.481 &3.552\textpm0.084 \\
\hline
Ours&0.627&0.091 &1.862 &4.074\textpm0.096\\
\hline
\end{tabular}}
\end{table}

\subsection{Evaluation on SpeechCodeVAE}

First, we evaluated the performance of speech content tokens extracted by SpeechCodeVAE, randomly selecting 100 utterances from the L2-ARCTIC corpus with three configurations to show the effectiveness of proposed SpeechCodeVAE: 1) CosyVoice-50Hz~\cite{cosyvoice} baseline, 2) SpeechCodeVAE without CTC loss, 3) Full SpeechCodeVAE. ~\autoref{tab:codec} shows our method achieves 59\% higher De-duplication Efficiency and 9× better Speed Robustness than baseline, demonstrating superior temporal robustness through our Content Encoder. In the ablation experiment on SpeechCodeVAE without CTC loss, we found that its performance was significantly worse than the other two models, which demonstrates that CTC loss plays a critical role in the continuity and robustness of tokens. 
We define the locality property as discrete tokens per frame encapsulating information solely from their corresponding speech segments while excluding cross-frame dependencies. Experimental results substantiate that our SpeechCodeVAE achieves superior locality compared with baseline models.
\begin{table}[th]
\centering
\caption{Objective Evaluation of SpeechCodeVAE Performance}
\label{tab:codec}
\scalebox{0.7}{\begin{tabular}{ccc}
\hline
&De-duplication Efficiency$\uparrow$ &Speed Robustness$\uparrow$ \\
\hline
CosyVoice-50Hz &0.159 &0.024 \\
\hline
w/o CTC&0.086 &0.009 \\
\hline
SpeechCodeVAE&\textbf{0.253}& \textbf{0.219} \\
\hline
\end{tabular}}
\end{table}
\begin{figure*}[t]
\centering
\includegraphics[width=0.96\linewidth]{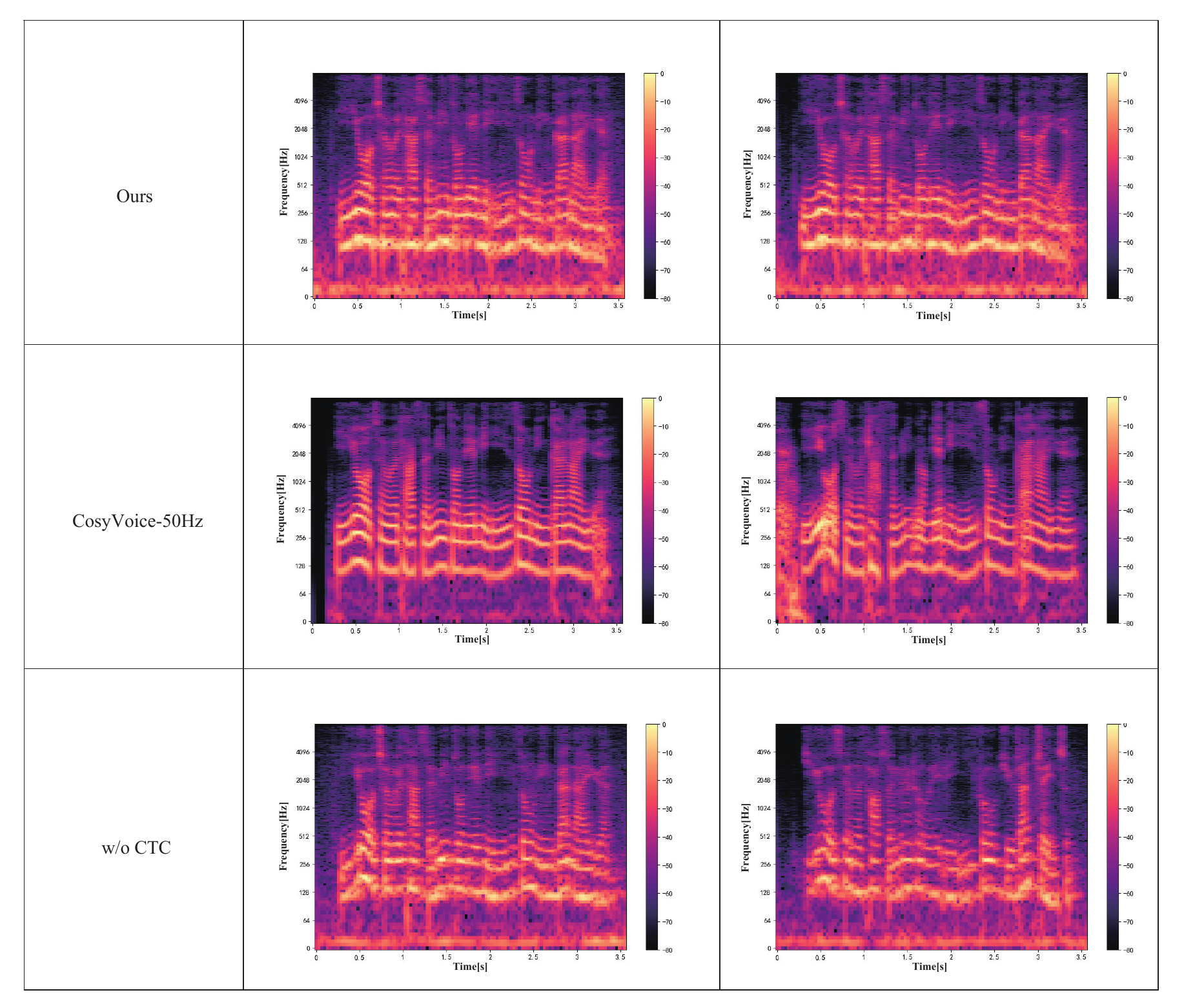} 
\caption{Comparative spectrogram analysis of odd-even token repetition replacement.}
\label{Fig: fig4}
\end{figure*}

Next, to visually demonstrate the strong locality of the tokens extracted by SpeechCodeVAE, we performed an evaluation via odd-even token repetition replacement: replaced the even-positioned tokens in discrete speech token sequence with the values of odd positions, then reconstructed speech waveforms with a unified timbre, and plotted spectrograms. This operation was performed on the three models above, and the results are shown in ~\autoref{Fig: fig4}. It can be observed that for SpeechCodeVAE, the spectrograms before and after token odd-even replacement are nearly identical, with no obvious changes in high and low frequencies. In contrast, for the Cosyvoice-50Hz model, high-frequency details in the spectrograms are notably lost, while the spectrograms of the model without CTC exhibits substantial changes in the low-frequency components. These results further highlight the superiority of SpeechCodeVAE in preserving spectral integrity and structural robustness across frequency domains, which verifies that the speech content tokens extracted by SpeechCodeVAE possess superior locality.

\begin{table}[th]
\centering
\caption{The VC Performance Evaluation of SpeechCodeVAE}
\label{tab:vc}
\scalebox{0.73}{\begin{tabular}{ccccccc}

\hline
&Sim-O$\uparrow$ &Sim-R$\uparrow$ &CMOS$\uparrow$ &SMOS$\uparrow$ \\
\hline
YourTTS&0.326 &0.474 &3.961\textpm 0.134 & 3.950\textpm 0.107 \\
\hline
FreeVC&0.282 &0.530 &3.810\textpm 0.122 &4.144\textpm 0.083\\
\hline
Ours-Kmeans& 0.314 &0.425 & 3.683\textpm 0.139  &4.091\textpm 0.147\\
\hline
Ours&\textbf{0.406}&\textbf{0.606}&\textbf{4.256\textpm0.118}&\textbf{4.302\textpm0.092}\\
\hline
\end{tabular}}
\end{table}


Finally, to evaluate the capability of the SpeechCodeVAE in disentangling and reconstructing speech content and timbre, we employed VC as the validation task. We utilized 50 utterances from the L2-ARCTIC corpus as source speech and randomly select 50 utterances from the same corpus as target references. This experimental design ensures both the source and target speaker utterances used in VC are not present during the training of SpeechCodeVAE, thereby providing a rigorous assessment of its generalization capabilities. For baseline comparisons, we selected YourTTS~\cite{yourtts} and Free VC~\cite{freevc}, both recognized for their superior voice conversion performance. Additionally, we implemented a comparative approach by replacing the VQ-based speech tokens quantization method with K-means clustering. The experimental results are presented in ~\autoref{tab:vc}.

Experimental results demonstrate that our SpeechCodeVAE significantly outperforms comparable VC models. This superiority is attributed to enhanced content modeling capabilities for linguistically unseen domains beyond the training data distribution. Notably, the K-means quantization approach exhibits degraded synthesis quality compared to VQ-based methods, manifesting as the decrease in signal-to-noise ratio (SNR). This degradation stems from codebook-parameter mismatch during inference, where the K-means derived codebook fails to align with the decoder's learned latent representations.

\subsection{TTS Results}
In this section, we evaluate the effectiveness of the auxiliary TTS task. Our testing method involves creating 100 English text sentences, randomly selecting 100 speech samples from the LibriSpeech test-clean as target speakers utterances, and then synthesizing speech. For the baseline models, we choose YourTTS~\cite{yourtts} and NaturalSpeech2 (NS2)~\cite{naturalspeech2}. Additionally, to verify the effectiveness of our SpeechRestorer (SR) and Variance Adapter (VA), we conducted ablation experiments for comparative verification: for SpeechRestorer, we simply removed this component during inference; for Variance Adapter, we retrained the model after removing the Variance Adapter. The experimental results are shown in ~\autoref{tab:tts}.
\begin{table}[th]
\centering
\caption{The TTS Performance Evaluation}
\label{tab:tts}
\scalebox{0.8}{\begin{tabular}{ccccc}
\hline
  &WER$\downarrow$ &Sim-O$\uparrow$ &CMOS$\uparrow$ &SMOS$\uparrow$ \\
\hline
YourTTS &0.120 &0.503 &3.786\textpm0.176 & 4.125\textpm0.162\\
NS2 &\textbf{0.063} &\textbf{0.652} & \textbf{3.944\textpm0.153} & \textbf{4.263\textpm0.095}\\
w/o SR &0.090 &0.625&3.629\textpm0.146 &4.156\textpm0.066\\

w/o VA &0.105 &0.594 &3.537\textpm0.151 &3.926\textpm0.075\\
\hline
Ours &0.084 &0.620 &3.850\textpm0.141 &4.204\textpm0.090\\
\hline
\end{tabular}}
\end{table}

While the TTS evaluation reveals performance gaps between our model and NaturalSpeech2 (CMOS: 3.85 vs. 3.94), this limitation is not solely due to the suboptimal generalization of the pretrained Speaker Encoder model, but primarily from the fact that NaturalSpeech2 is trained on nearly two orders of magnitude more data than our approach. This vast difference in training scale fundamentally enhances NaturalSpeech2's ability to model intricate acoustic-phonetic patterns and cross-lingual variations.

Additionally, removing SpeechRestorer led to a decrease in the CMOS evaluation score. Our proposed SpeechRestorer demonstrates remarkable efficacy in enhancing synthetic speech fluency through its novel error-correction mechanism that compensates for acoustic discontinuities in the decoding pipeline. After removing the Variance Adapter, all metrics significantly declined. This indicates that prosodic information, which is primarily modeled by the Variance Adapter, exerts critical influence on all aspects of speech synthesis.

\section{Conclusion}
We propose a novel framework, SpeechAccentLLM, to address the challenges of FAC within the paradigm of LLMs. During the content discretization process of the SpeechCodeVAE model, we innovatively introduce a CTC-guided codebook discretization structure. The extracted content tokens exhibit strong locality, which is well-supported by both objective metrics and visualization results. The reconstruction capability of SpeechCodeVAE, including timbre reconstruction and speech generation, is validated through VC experiments.

Our joint training strategy, which integrates FAC and TTS tasks by leveraging the extracted speech content tokens, offers effective solutions to persistent challenges in FAC research, such as data scarcity and convergence instability. Furthermore, to alleviate occasional lexical errors and improve acoustic incoherence when using LLMs for speech processing, we propose SpeechRestorer,which uses locality of speech tokens to introduce a new research approach in this field.

However, the model still has certain limitations. For example, the prosody modeling is not sufficiently comprehensive, the timbre reconstruction effect is affected by the pre-trained model, and SpeechRestorer cannot improve the phenomena of skipped words and repetitions. In future work, our aim is to explore more robust speech discrete tokens to enhance three key dimensions under the LLM paradigm: (1) enhancing model generalization capability, (2) improving speech synthesis quality, and (3) increasing timbre similarity.



\bibliography{custom}

\end{document}